\documentclass[12pt]{article}
\usepackage[dvipdfmx]{graphicx}

\usepackage{amsmath,amssymb,amscd}
\usepackage{latexsym}
\usepackage{bm}
\usepackage{booktabs}
\usepackage{mathrsfs}
\usepackage{float}
\usepackage{fancybox}
\usepackage{braket}
\usepackage{ascmac}
\usepackage{fancybox,pifont,layout}
\usepackage{amsmath,amsthm}
\usepackage{mediabb}
\usepackage{cite}
\usepackage{url}
\usepackage{arydshln}%破線のパッケージ
\newcommand{\al}[1]{\begin{align}#1\end{align}} %%\al の定義
\newcommand{\mat}[1]{\left( \begin{matrix}#1\end{matrix} \right)}%行列簡略化

\usepackage{color}
\usepackage[square,sort&compress,comma,numbers]{natbib}
\renewcommand{\thefootnote}{(\roman{footnote})}

\newcommand{\cwhite}{\textcolor{white}}

\makeatletter

\@addtoreset{equation}{section}
\makeatother

\usepackage{hyperref}
\hypersetup{
	colorlinks=true,
	linkcolor=red,
	citecolor=blue
}
\begin{document}
\baselineskip=18pt
%%%%%%%%%%%%%%%%%%%%%%%%%%%%%%%%%%%%%%%%%%%%%%%%%%%%%%%%%%%%%%%%%
\begin{titlepage}
%%%%% PREPRINT NUMBERS %%%%%%
\begin{flushright}
{KOBE-TH-19-03}\\
\end{flushright}
\vspace{1cm}
%%%%%%%%%%%%%%%% TITLE %%%%%%%%%%%%%%%%%%%%%%%%%%%%%%%%%%%%%%%%%%
\begin{center}{\Large\bf 
5d Dirac fermion on quantum graph
}
\end{center}
%%%%%%%%%%%%%%%% AUTHORS %%%%%%%%%%%%%%%%%%%%%%%%%%%%%%%%%%%%%%%%
\vspace{0.5cm}
\begin{center}
Yukihiro Fujimoto$^{(a)}$\footnote{E-mail: \url{y-fujimoto@oita-ct.ac.jp}},
Tomonori Inoue$^{(b)}$\footnote{E-mail: {\url{186s102s@stu.kobe-u.ac.jp}}},
Makoto Sakamoto$^{(b)}$\footnote{E-mail: \url{dragon@kobe-u.ac.jp}},\\
Kazunori Takenaga$^{(c)}$\footnote{E-mail: \url{takenaga@kumamoto-hsu.ac.jp}} and
Inori Ueba$^{(b)}$\footnote{E-mail: \url{i-ueba@stu.kobe-u.ac.jp}}
\end{center}
%%%%%%%%%%%%%%%%%%%%%%% AFFILIATION %%%%%%%%%%%%%%%%%%%%%%%%%%%%%
\begin{center}
%\small
${}^{(a)}${\it National Institute of Technology, Oita college, \\
Maki1666, Oaza, Oita 870-0152, Japan}\\
${}^{(b)}${\it Department of Physics, Kobe University, 
Kobe 657-8501, Japan}\\
{${}^{(c)}${\it {Faculty of Health Science}, Kumamoto Health Science University\\
325 Izumi-machi, Kita-ku, Kumamoto 861-5598, Japan }}\\

%%%%%
%%%%%%%
\end{center}

%%%%%%%%%%%\UTF{0081}@\UTF{0081}@\UTF{0083}A\UTF{0083}u\UTF{0083}X\UTF{0083}g\UTF{0083}\UTF{0089}\UTF{0083}N\UTF{0083}g\UTF{0081}@\UTF{0081}@%%%%%%%%%%%
\vspace{1cm}
\begin{abstract}
In this paper, we investigate a five-dimensional Dirac fermion on a quantum graph that consists of a single vertex and {$N$} loops. We find that the model possesses a rich structure of boundary conditions {for wavefunctions on the quantum graph} and they can be classified into $(2N+1)$ distinct categories. It is then shown that there appear degenerate four-dimensional chiral massless fermions in the four-dimensional mass spectrum. We briefly discuss how our model could naturally solve the problems of the fermion generation, the fermion mass hierarchy and the origin of the \textit{CP}-violating phase.
\end{abstract}
\end{titlepage}
	%%%%%%%%%%%%%%%%%%%%%%%%%%%%%%%%%%%%%%%%
\newpage

\setcounter{footnote}{0}
\renewcommand{\thefootnote}{\arabic{footnote}\,}
%%%%%%%%%%%%\UTF{0081}	section.1\UTF{0081}	%%%%%%%%%%%%%
\section{Introduction}
\quad The standard model (SM) has been completed by the discovery of the Higgs boson. However, the SM still contains several mysteries and problems, which will remain to be solved. One is so-called the generation problem. The SM includes three sets of quarks and leptons, which have exactly the same quantum numbers except for the Yukawa couplings. Why nature provides the three generation of quarks and leptons is veiled. Another problem is the fermion mass hierarchy. Even though the second and the third generations of the quarks and the charged leptons are just copies of the first generation, their masses have an exponential hierarchy around $10^{5}$. The third problem is the origin of the \textit{CP} violating phase in the Cabbibo-Kobayashi-Maskawa (CKM) matrix.

There are many attempts to try to solve the problems mentioned above. One of promising candidates will be higher-dimensional theory on extra dimensions. If extra dimensional models give the degeneracy of four-dimensional (4d) massless chiral fermions in 4d mass spectrum of Kaluza-Klein (KK) decomposition, the degeneracy will explain the {generation} of the quarks and the leptons.{\footnote{{A fascinating way to produce the three-generation in 4d mass spectrum is to introduce a vortex background~\cite{Libanov:2000uf,Frere:2000dc} or a magnetic flux~\cite{Blumenhagen:2000wh,Blumenhagen:2000ea,Cremades:2004wa,Abe:2008fi,Abe:2008sx,Abe:2013bca,Fujimoto:2013xha,Abe:2014noa,Abe:2015yva} in extra dimensions. Another promising way is to generalize boundary conditions on extra dimensions~\cite{Fujimoto:2016llj,Fujimoto:2016rbr,Fujimoto:2018cnf,Fujimoto:2018tjm}.}}} The fermion mass hierarchy could be resolved if the zero mode profiles are localized near some points on extra dimensions. This is because the mass matrix $m_{ij}$ between the \textit{i}-th generation and the \textit{j}-th one of the fermions will be of the form
	\al{
	m_{ij}=g_{Y}v\int dy \bigl(f^{(i)}_{0}(y)\bigr)^{\ast}{g^{(j)}_{0}(y)},\label{mass-matrix}
	}
where $g_{Y}$ is a Yukawa coupling constant\footnote{It should be emphasized that the Yukawa coupling $g_{Y}$ has no index with respect to the generation. This is because we are considering extra dimensional models that could solve the generation problem, so that the models {should} be assumed to contain only one generation of the quarks and the lepton fields.}, {$v$ is a vacuum expectation value of the Higgs field, $\int dy$ denotes the integral over the extra dimensions,} and $f_{0}^{(i)}(y)$ ($g_{0}^{(j)}(y)$) is the profile of the \textit{i}-th (\textit{j}-th) generation of the 4d massless chiral fermions. Then, if $f_{0}^{(i)}(y)$ and $g_{0}^{(j)}(y)$ are localized functions on the extra dimensions, we will get a hierarchical mass matrix, which is sensitive to locations of localization~{\cite{ArkaniHamed:1998rs,ArkaniHamed:1999dc,Dvali:2000ha,Gherghetta:2000qt,Huber:2000ie,Kaplan:2001ga,Fujimoto:2011kf}}. A \textit{CP}-violating phase can appear if the zero mode functions $f_{0}^{(i)}(y)$ and $g_{0}^{(j)}(y)$ are genuine complex functions, so that the mass matrix (\ref{mass-matrix}) can possess complex phases naturally~{\cite{Fujimoto:2013ki}}.

\begin{figure}[htbp]
		\begin{center}
		\begin{minipage}{0.4\hsize}
		\begin{center}
		\includegraphics[width=45mm]{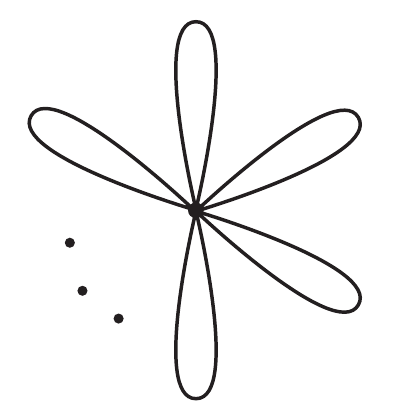}
		\par
		\vspace{0.0cm}
		\caption{Rose graph with one vertex and $N$ loops}
		\label{fig:rose-graph}
		\end{center}
		\end{minipage}
		\hspace{2cm}
		\begin{minipage}{0.4\hsize}
		\begin{center}
		\includegraphics[width=45mm]{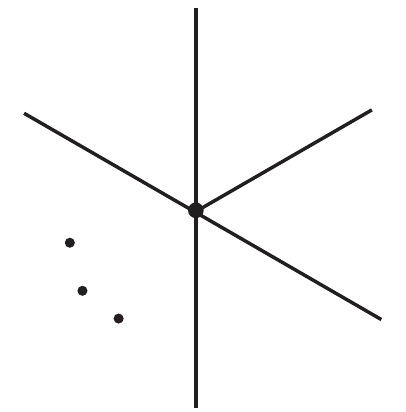}
		\par
		\vspace{0.0cm}
		\caption{Star graph with one vertex and $N$ bonds}
		\label{fig:star-graph}
		\end{center}
		\end{minipage}
		\end{center}
		\vspace{1cm}
		\begin{center}
		\begin{minipage}{0.6\hsize}
		\begin{center}
		\includegraphics[width=50mm]{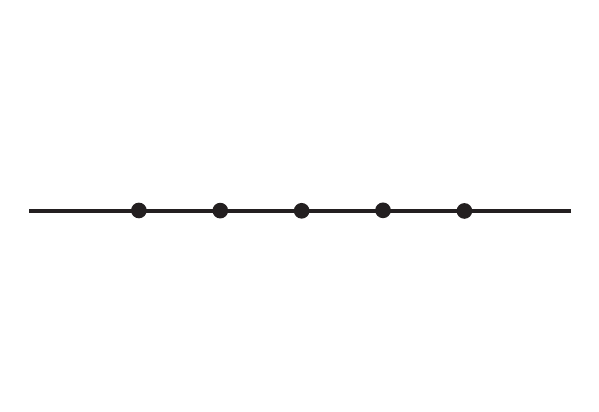}
		\par
		\vspace{-1.5cm}
		\caption{Interval with {$(N-1)$} point interactions}
		\label{fig:interval}
		\end{center}
		\end{minipage}
		\end{center}
%%%%%%%%%%%%%%%%%%%%%%%%%%%%%%%%%%%%%%%%%%%%%%%%
%		\begin{minipage}{0.47\textwidth}
%		\begin{center}
%		\scalebox{0.8}{\includegraphics{figure/Fig_1c.eps}}
%		\par
%		\vspace{0.0cm}
%		{\footnotesize{(c) $L_{+}L_{-}<0$, $L_{+}+L_{-} \le 0$}}
%		\end{center}
%		\end{minipage}
%		\hfill
%		\begin{minipage}{0.47\textwidth}
%		\begin{center}
%		\scalebox{0.8}{\includegraphics{figure/Fig_1d.eps}}
%		\par
%		\vspace{0.0cm}
%		{\footnotesize{(d) $L_{+}L_{-}\ge 0$, $L_{+}+L_{-}\le 0$}}
%		\end{center}
%		\end{minipage}
\end{figure}

In order to {obtain} extra dimensional models that possess the properties mentioned in the above paragraph, we investigate a five-dimensional (5d) Dirac action on a quantum graph, which consists of bonds and vertices. (For reviews of quantum graph, see {\cite{Kuchment_2004,Kuchment_2005}}.) In this paper, we take, as a quantum graph, a rose graph consisting of one vertex and $N$ bonds, where each bond forms a loop that begins and ends at the vertex (see fig.~\ref{fig:rose-graph}).

%%%%%%%%%%%%%%% Figure 1 %%%%%%%%%%%%%%%%%%%%%%
%%%%%%%%%%%%%%%%%%%%%%%%%%%%%%%%%%%%%%%%%%%%%%%
%%%%%%%%%%%%%%%%%%%%%%%%%%%%%%%%%%%%%%%%%%%%%%%

One of the reason to consider the rose graph is that ordinary one-dimensional extra dimensions, like a circle or an interval, cannot solve the generation problem because no desired degeneracy appears in the spectrum. Another reason is that the rose graph with $N$ loops may be regarded as a master quantum graph with $N$ bonds because the rose graph will reduce to so-called {a} star graph\footnote{{There are many studies in terms of star graphs, {see \textit{e.g.}}~\cite{Berkolaiko_1999,Bellazzini:2008mn,Adami:2010zj,Fujimoto:2018lzq}.} In particular, several works have been carried out from the viewpoint of extra dimensions on star graphs which are realized by flux compactifications in string theory,  see~\cite{Dimopoulos:2001qd,Kim:2005aa,Cacciapaglia:2006tg,Bechinger:2009qk,Abel:2010kw,Law:2010pv}.} (see fig.~\ref{fig:star-graph}), an interval with $(N-1)$ point interactions{\footnote{{One-dimensional quantum mechanics with point interactions has attractive properties~\cite{Nagasawa:2002un,Nagasawa:2003tw,Nagasawa:2005kv} and is applied to extra-dimension models to solve the generation problem and the fermion mass hierarchy~\cite{Fujimoto:2012wv,Fujimoto:2014fka,Fujimoto:2017lln}.}}} (see fig.~\ref{fig:interval}) and so on, by appropriately tuning boundary conditions (or connection conditions) {for wavefunctions} at the vertex of the rose graph.

In this paper, we show that allowed boundary conditions (BCs) on the rose graph can be classified into $(2N+1)$ distinct types of BCs, and clarify how many massless chiral fermions appear in the 4d mass spectrum for each type of BCs. The results show that our model possesses the desired properties mentioned in the second paragraph in this section.

The paper is organized as follows: In the next section, we give a setup of our model. In Section 3, we classify allowed boundary conditions into $(2N+1)$-types of them. In Section 4, we examine zero mode solutions for each type of the boundary conditions. In Section 5, we discuss {a topological nature of the zero modes}. The Section 6 is devoted to conclusion and discussion.

%%%%%%%%%%%%\UTF{0081}	section.2\UTF{0081}	%%%%%%%%%%%%%
\section{5d Dirac action on quantum graph}
\quad The 5d Dirac action we consider is given by
	\al{
	S=\int d^{4}x\sum^{N}_{a=1}\int_{L_{a-1}}^{L_{a}}dy\,\overline{\Psi}(x,y)[i\gamma^{\mu}\partial_{\mu}+i\gamma^{y}\partial_{y}{+M}]\Psi(x,y),\label{action}
	}
where $x^{\mu}$ ($\mu=0,1,2,3$) denote the coordinates of the 4d Minkowski space-time and $y$ is the coordinate of an extra dimension. $\Psi(x,y)$ is a four-component Dirac spinor on five dimensions and the Dirac conjugate {$\overline{\Psi}$} is defined by $\overline{\Psi}=\Psi^{\dagger}\gamma^{0}$. $\gamma^{\mu}$ ($\mu=0,1,2,3$) are $4\times 4$ gamma matrices and $\gamma^{y}$ is taken to be $\gamma^{y}=-i\gamma^{5}$ ($\gamma^{5}=i\gamma^{0}\gamma^{1}\gamma^{2}\gamma^{3}$). $M$ is the bulk mass of the 5d Dirac fermion.

\begin{figure}[ht]
	\begin{center}
	\includegraphics[width=80mm]{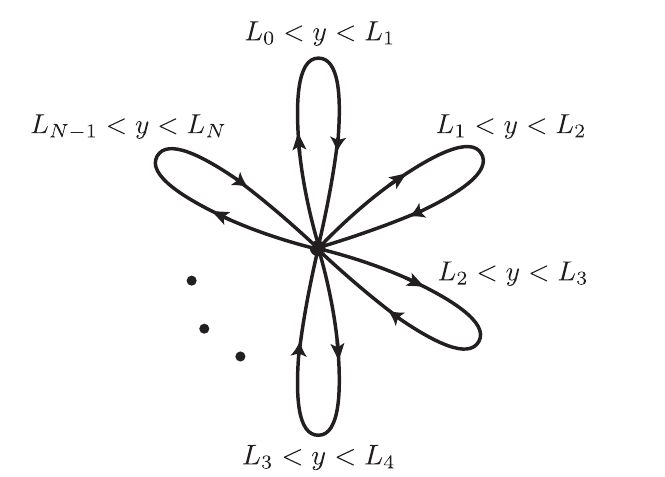}
	\end{center}
	\vspace{-0.3cm}
	\caption{A rose graph consisting of one vertex and $N$ loops.}
	\label{fig:rose_graph}	
	\end{figure}

In this paper, we consider, as the extra dimension, a rose graph consisting of one vertex and {$N$} bonds, each of which forms a loop with $L_{a-1}<y<L_{a}$ ($a=1,2,\cdots,N$): see fig.~\ref{fig:rose_graph}.
	%%%%%%%%%%%%%	figure	%%%%%%%%%%
	%%%%%%%%%%%%%%%%%%%%%%%%%%%%%%%%%%
	
The action principle $\delta S=0$ leads to the 5d Dirac equation
	\al{
	[i\gamma^{\mu}\partial_{\mu}+i\gamma^{y}\partial_{y}{+M}]\Psi(x,y)=0,\label{eom}
	}
together with 
	\al{
	\sum^{N}_{a=1}\bigl[\overline{\Psi}(x,y)\gamma^{5}\delta \Psi(x,y)\bigr]^{y=L_{a}-\varepsilon}_{y=L_{a-1}+\varepsilon}=0, \label{surface-term}
	}
where $\varepsilon$ is an infinitesimal positive constant. The condition (\ref{surface-term}) may be understood as the momentum conservation or the probability current conservation in the direction of the extra dimension at the center of the rose graph depicted in fig.~\ref{fig:rose_graph}. As we {will} see in the next section, eq.~(\ref{surface-term}) leads to boundary conditions that the {field} $\Psi(x,y)$ should obey at the boundaries $y=L_{0}+\varepsilon, L_{1}\pm\varepsilon,\cdots,L_{N-1}\pm\varepsilon, L_{N}-\varepsilon$.

In terms of the {4d right-handed (left-handed)} chiral spinors {$\psi^{(i)}_{\textrm{R},n}(x)$ $\bigl(\psi^{(i)}_{{\rm L},n}(x)\bigr)$}, the 5d Dirac spinor $\Psi (x,y)$ will be expanded into the form
	\al{
	\Psi(x,y)=\sum_{i}\sum_{n}\psi^{(i)}_{{\rm R},n}(x)f^{(i)}_{n}(y)+\sum_{i}\sum_{n}\psi^{(i)}_{{\rm L},n}(x)g^{(i)}_{n}(y),\label{KK-expansion}
	}
where the index $n$ indicates the \textit{n}-th level of the KK modes and \textit{i} denotes the index that distinguishes the degeneracy of the \textit{n}-th KK modes (if it exists). The mode functions $f^{(i)}_{n}(y)$ and $g^{(i)}_{n}(y)$ are assumed to form a complete set with respect to the extra dimensional space and satisfy the orthonormality relations
	\al{
	\sum^{N}_{a=1}\int^{L_{a}}_{L_{a-1}}dy\,\bigl(f^{(i)}_{n}(y)\bigr)^{\ast}f^{(i')}_{n'}(y)=\delta_{nn'}\delta^{ii'},\label{orthonormality-relation-f}\\
	\sum^{N}_{a=1}\int^{L_{a}}_{L_{a-1}}dy\,\bigl(g^{(i)}_{n}(y)\bigr)^{\ast}g^{(i')}_{n'}(y)=\delta_{nn'}\delta^{ii'}.\label{orthonormality-relation-g}
	}
	
Substituting the expansion (\ref{KK-expansion}) into eq.~(\ref{eom}) and using the relations {$i\gamma^{\mu}\partial_{\mu}\psi^{(i)}_{{\rm R},n}(x)+m_{n}\psi^{(i)}_{{\rm L},n}(x)=0$ and $i\gamma^{\mu}\partial_{\mu}\psi^{(i)}_{{\rm L},n}(x)+m_{n}\psi^{(i)}_{{\rm R},n}(x)=0$}, which are the {equations} of motion for the 4d chiral spinors $\psi^{(i)}_{{{\rm R}/{\rm L}},n}(x)$ with mass $m_{n}$, we find the equations that the mode functions $f^{(i)}_{n}(y)$ and $g^{(i)}_{n}(y)$ should satisfy
	\al{
	(\partial_{y}+M)f^{(i)}_{n}(y)=m_{n}g^{(i)}_{n}(y),\label{SUSYQM-f}\\
	(-\partial_{y}+M)g^{(i)}_{n}(y)=m_{n}f_{n}^{(i)}(y).\label{SUSYQM-g}
	}
It follows that $f^{(i)}_{n}(y)$ and $g^{(i)}_{n}(y)$ satisfy the eigenvalue equations
	\al{
	(-\partial_{y}^{2}+M^{2})f^{(i)}_{n}(y)=m_{n}^{2}f^{(i)}_{n}(y),\\
	(-\partial_{y}^{2}+M^{2})g^{(i)}_{n}(y)=m_{n}^{2}g^{(i)}_{n}(y).
	}
Then, inserting the expansion (\ref{KK-expansion}) into the action (\ref{action}), and using the relations (\ref{orthonormality-relation-f}) - (\ref{SUSYQM-g}), we can rewrite the action $S$, in terms of the 4d spinors, as
	\al{
	S&=\int d^{4}x\Bigl\{\sum_{i}\overline{\psi}^{(i)}_{{\rm R},0}(x) i\gamma^{\mu}\partial_{\mu}\psi^{(i)}_{{\rm R},0}(x)+\sum_{{j}}\overline{\psi}^{({j})}_{{\rm L},0}(x) i\gamma^{\mu}\partial_{\mu}\psi^{({j})}_{{\rm L},0}(x)\nonumber\\
	&\hspace{2cm}+\sum_{i}\sum_{n\neq 0}\overline{\psi}^{(i)}_{n}(x) (i\gamma^{\mu}{\partial_{\mu}}{+m_{n}})\psi^{(i)}_{n}(x)\Bigr\},
	}
where $\psi^{(i)}_{n}(x)\equiv \psi^{(i)}_{{\rm R},n}(x)+\psi^{(i)}_{{\rm L},n}(x)$ for $n\neq 0$, and $\psi^{(i)}_{{\rm R},0}(x)$, $\psi^{(i)}_{{\rm L},0}(x)$ denote the 4d chiral massless spinors with $m_{0}=0$.

%%%%%%%%%%%%\UTF{0081}	section.3	%%%%%%%%%%%%%
\section{Classification of boundary conditions}
\quad In the previous section, we have succeeded in expressing the action $S$ in terms of the 4d mass eigenmodes. However, in order to determine the mass eigenvalue $m_{n}$ as well as the eigenfunctions {$f^{(i)}_{n}(y)$} and $g^{(i)}_{n}(y)$, we need to specify some boundary conditions for the mode functions {$f^{(i)}_{n}(y)$} and $g^{(i)}_{n}(y)$ at the boundaries $y=L_{0}+\varepsilon, L_{1}\pm\varepsilon,\cdots,L_{N-1}\pm\varepsilon, L_{N}-\varepsilon$. In this section, we derive allowed boundary conditions for {$f^{(i)}_{n}(y)$} and $g^{(i)}_{n}(y)$ from eq.~(\ref{surface-term}) and {classify them}.

Substituting the expansion (\ref{KK-expansion}) of $\Psi(x,y)$ (and also $\delta\Psi(x,y)$) into eq.~(\ref{surface-term}) and using the fact that $\psi^{(i)}_{{\rm R}/{\rm L},n}(x)$ and $\delta \psi^{(i)}_{{\rm R}/{\rm L},n}(x)$ are independent fields, we find
	\al{
	\vec{F}^{(i)}_{n}{}^{\dagger}\vec{G}^{(j)}_{m}=\vec{G}^{(j)}_{m}{}^{\dagger}\vec{F}^{(i)}_{n}=0\qquad \text{for}\quad {}^{\forall}n,m\ \text{and}\ {{}^{\forall}i,j},\label{orthogonal-condition}
	}
where $\vec{F}^{(i)}_{n}$ and $\vec{G}^{(j)}_{m}$ are $2N$-dimensional complex vectors defined by
	\al{
	\vec{F}^{(i)}_{n}\equiv \mat{f^{(i)}_{n}(L_{0}+\varepsilon)\\ f^{(i)}_{n}(L_{1}-\varepsilon)\\ f^{(i)}_{n}(L_{1}+\varepsilon)\\ f^{(i)}_{n}(L_{2}-\varepsilon)\\ \vdots \\ f^{(i)}_{n}(L_{a-1}+\varepsilon)\\ f^{(i)}_{n}(L_{a}-\varepsilon)\\ \vdots \\ f^{(i)}_{n}(L_{N-1}+\varepsilon) \\ f^{(i)}_{n}(L_{N}-\varepsilon)},\qquad {\vec{G}^{(j)}_{m}\equiv \mat{g^{(j)}_{m}(L_{0}+\varepsilon)\\ -g^{(j)}_{m}(L_{1}-\varepsilon)\\ g^{(j)}_{m}(L_{1}+\varepsilon)\\ -g^{(j)}_{m}(L_{2}-\varepsilon)\\ \vdots \\ g^{(j)}_{m}(L_{a-1}+\varepsilon)\\ -g^{(j)}_{m}(L_{a}-\varepsilon)\\ \vdots \\ g^{(j)}_{m}(L_{N-1}+\varepsilon) \\ -g^{(j)}_{m}(L_{N}-\varepsilon)}}.\label{boundary-vectors}
	}
Here, we may call $\vec{F}^{(i)}_{n}$ and $\vec{G}^{(j)}_{m}$ boundary vectors.

Let $V$ $(W)$ be the vector space spanned by $\{\vec{F}^{(i)}_{n}\}$ ($\{\vec{G}^{(j)}_{m}\}$). Then, $V$ is found to be orthogonal to $W$ in a sense of eq.~(\ref{orthogonal-condition}) and will be regarded as the orthogonal complement space of $W$. It follows from this observation that eq.~(\ref{orthogonal-condition}) {may} be replaced by the condition
	\al{
	\mathcal{P}_{-}\vec{F}^{(i)}_{n}=\mathcal{P}_{+}\vec{G}^{(j)}_{m}=0\qquad \text{for}\quad {}^{\forall}n,m\ \text{and}\ {{}^{\forall}i,j,}
	}
where $\mathcal{P}_{\pm}$ are projection matrices that $\mathcal{P_{+}}$ ($\mathcal{P_{-}}$) maps the $2N$-dimensional complex vector space into $V$ ($W$), and they are assumed to satisfy $\mathcal{P_{+}}+\mathcal{P_{-}}=I_{2N}$ ($I_{2N}$ is the $2N\times 2N$ identity matrix), $(\mathcal{P_{\pm}})^{2}=\mathcal{P_{\pm}}$, $\mathcal{P_{+}}\mathcal{P_{-}}=\mathcal{P_{-}}\mathcal{P_{+}}=0$ and $(\mathcal{P_{\pm}})^{\dagger}=\mathcal{P_{\pm}}$.

The projection matrices $\mathcal{P_{\pm}}$ can be {represented} as
	\al{
	\mathcal{P_{\pm}}=\frac{1}{2}(I_{2N}\pm U),
	}
where $U$ is a $2N\times 2N$ Hermitian matrix with the property
	\al{
	U^{2}=I_{2N}.\label{U2-condition}
	}
Thus, we found that the mode functions $f^{(i)}_{n}(y)$ and {$g^{(j)}_{m}(y)$} obey the following boundary conditions\footnote{It will be {worthwhile pointing} out that for one-dimensional quantum mechanics on a quantum graph boundary conditions are generally imposed on not only $\psi(y)$ but also $\psi'(y)=\frac{d\psi(y)}{dy}$~\cite{Kostrykin:1998gz,Fulop:1999pf}, though our model gives conditions only for the wavefunctions $f^{(i)}_{n}(y)$ and $g^{(i)}_{n}(y)$ but not $f'{}^{(i)}_{n}(y)$ and $g'{}^{(i)}_{n}(y)$.}
	\al{
	(I_{2N}-U)\vec{F}^{(i)}_{n}={\vec{0}},\label{BC-F}\\
	(I_{2N}+U){\vec{G}^{(j)}_{m}}={\vec{0}},\label{BC-G}
	}
for {${}^{\forall}n,m$ and ${}^{\forall}i,j$}. Thus, we conclude that a Hermitian matrix $U$ satisfying eq.~(\ref{U2-condition}) specifies a 5d Dirac theory on a rose graph depicted in {fig.~\ref{fig:rose_graph}.}

For convenience of later analysis, we classify eqs.~(\ref{BC-F}) {and} (\ref{BC-G}) into the $(2N+1)$ types of the boundary conditions. Since $U^{2}=I_{2N}$, the eigenvalues of $U$ are $+1$ or $-1$, so that the matrix $U$ can be classified by the number of the eigenvalue $+1$ (or $-1$). We then call the boundary condition the {type\,($2N-k$,$k$)} if the number of the eigenvalue $+1$ ($-1$) of $U$ is $2N-k$ ($k$) for $k=0,1,2,\cdots,2N$, so that for the {type\,($2N-k$,$k$)} BC the matrix $U$ can be expressed as
	\al{
	\text{type}\,(2N-k,k):\qquad U&=V\left(\begin{array}{c:c}
						\begin{array}{ccc}
						+1& &0\\
						&\ddots&\\
						0& &+1
						\end{array}&\begin{array}{ccc}
						\cwhite{0}\cwhite{0}&\cwhite{0}\\
						\cwhite{0}&0&\cwhite{0}\\
						\cwhite{0}&\cwhite{0}&\cwhite{0}
						\end{array}\\
						\hdashline
						\begin{array}{ccc}
						\cwhite{0}\cwhite{0}&\cwhite{0}\\
						\cwhite{0}&0&\cwhite{0}\\
						\cwhite{0}&\cwhite{0}&\cwhite{0}
						\end{array}&\begin{array}{ccc}
						-1& &0\\
						&\ddots&\\
						0& &-1
						\end{array}
					\end{array}\right)V^{\dagger},\label{type2N-kkBC}\\
					&\hspace{1.5cm}{\underbrace{\cwhite{testtesttesttes}}_{2N-k}\hspace{0.3cm}\underbrace{\cwhite{testtesttesttes}}_{k}}\nonumber
	}
for $V\in U(2N)$. Therefore the allowed boundary conditions on the rose graph are found to be classified into the $(2N+1)$-types of them. It follows from the form of (\ref{type2N-kkBC}) that the parameter space of the {type\,($2N-k$,$k$)} BC is {found} to be given by the coset space {$U(2N)/(U(2N-k)\times U(k))$}. It should be noticed that the parameter space of the {type\,($2N-k$,$k$)} BC is distinct from that of the {type\,($2N-k'$,$k'$)} BC for $k \neq k'$ and they are not continuously connected each other.

%%%%%%%%%%%%\UTF{0081}	section.4	%%%%%%%%%%%%%
\section{Zero mode solutions and boundary conditions}
\qquad In this paper, we restrict our {considerations} to the zero mode solutions $f^{(i)}_{0}(y)$ and $g^{(i)}_{0}(y)$, which obey the {equations} (see. eqs.~(\ref{SUSYQM-f}) {and} (\ref{SUSYQM-g}))
	\al{
	(\partial_{y}+M)f^{(i)}_{0}(y)=0,\label{zeromode-f}\\
	(-\partial_{y}+M){g^{(j)}_{0}(y)}=0.\label{zeromode-g}
	}
Any solutions to (\ref{zeromode-f}) {and} (\ref{zeromode-g}) will be written into the form
	\al{
	f^{(i)}_{0}(y)&{=}\sum^{N}_{a=1}\theta(y-L_{a-1})\theta(L_{a}-y)A^{(i)}_{a}C_{a}e^{-My},\label{f0}\\
	g^{(j)}_{0}(y)&{=}\sum^{N}_{a=1}\theta(y-L_{a-1})\theta(L_{a}-y)B^{(j)}_{a}C'_{a}e^{My}\label{g0}
	}
for ${}^{\exists}A_{a}^{(i)}, B_{a}^{(j)}\in \mathbb{C}$ ($a=1,2,\cdots, N$). The constants $C_{a}$ and $C'_{a}$, which are independent of $i$ and $j$, will be determined later {for our convenience}. The coefficients $A^{(i)}_{a}$ and $B^{(j)}_{a}$ should be chosen to satisfy the boundary conditions (\ref{BC-F}) {and} (\ref{BC-G}). Here, $\theta (y)$ is the Heaviside step function defined as $\theta (y)=1$ for $y>0$ and $\theta (y)=0$ for $y<0$. It should be emphasized that since we are considering functions on the rose graph depicted in {fig.~\ref{fig:rose_graph}}, we allow the mode functions $f^{(i)}_{n}(y)$ and $g^{(i)}_{n}(y)$ to be discontinuous at the boundaries $y=L_{0},L_{1},\cdots,L_{N}$.

The indices $i$ and $j$ of $f^{(i)}_{0}(y)$ and $g^{(j)}_{0}(y)$ denote the degeneracy of the zero mode solutions. The criterion that $K$ solutions $f^{(i)}_{0}(y)$ ($i=1,2,\cdots, K$) can be independent is that {the} $N$-dimensional complex vectors\footnote{Throughout this paper, we use the notation that $\vec{V}$ with the vector symbol {``\ $\vec{}$\,\ ''} denotes a $2N$-dimensional vector and the bold symbol $\mbox{\boldmath $A$}$ denotes an $N$-dimensional one.} $\mbox{\boldmath $A$}^{(i)}\equiv (A^{(i)}_{1},A^{(i)}_{2},\cdots, A^{(i)}_{N})^{\rm T}$ {($i=1,2,\cdots, K$)} associated with the solutions (\ref{f0}) are linearly independent. Similarly, $K'$ solutions $g^{(j)}_{0}(y)$ ($j=1,2,\cdots,K'$) can be independent if the $N$ dimensional complex vectors $\mbox{\boldmath $B$}^{(j)}\equiv (B^{(j)}_{1},B^{(j)}_{2},\cdots, B^{(j)}_{N})^{\rm T}$ {($j=1,2,\cdots,K'$)} associated with the solutions (\ref{g0}) are linearly independent. The above criterion will be used later.

In the following analysis, it will be convenient to introduce the vectors $\vec{\mathcal{F}}_{a}$ and $\vec{\mathcal{G}}_{a}$ ($a=1,2,\cdots,N$) which are the $2N$-dimensional vectors defined by
	\al{
	\vec{\mathcal{F}}_{a}\equiv C_{a}(\underbrace{0,\cdots,0}_{{2(a-1)}}, e^{-M(L_{a-1}+\varepsilon)},e^{-M(L_{a}-\varepsilon)},0,\cdots,0)^{\rm T},\\
	\vec{\mathcal{G}}_{a}\equiv C'_{a}(\underbrace{0,\cdots,0}_{{2(a-1)}}, e^{M(L_{a-1}+\varepsilon)},-e^{M(L_{a}-\varepsilon)},0,\cdots,0)^{\rm T},
	}
where the constants $C_{a}$ and $C'_{a}$ ($a=1,2,\cdots, N$) are the same as those given in eqs.~(\ref{f0}) and (\ref{g0}), and are chosen to be
	\al{
	\vec{\mathcal{F}}_{a}^{\dagger}\vec{\mathcal{F}}_{b}=\vec{\mathcal{G}}_{a}^{\dagger}\vec{\mathcal{G}}_{b}=\delta_{ab}\qquad (a,b=1,2,\cdots,N).\label{FFGG-normalization}
	}
{The important {observations}} are that $\vec{\mathcal{F}}_{a}$ are orthogonal to $\vec{\mathcal{G}}_{b}$, {\textit{i.e.}}
	\al{
	\vec{\mathcal{F}}_{a}^{\dagger}\vec{\mathcal{G}}_{b}=\vec{\mathcal{G}}_{a}^{\dagger}\vec{\mathcal{F}}_{b}=0\qquad (a,b=1,2,\cdots,N{)}\label{FFGG-orthogonal-relation}
	}
and furthermore that the set of $\{\vec{\mathcal{F}}_{a},\vec{\mathcal{G}}_{a}\ (a,b=1,2,\cdots,N)\}$ can be regarded as an {orthonormal basis} of the $2N$-dimensional complex vector space, so that any $2N$-dimensional complex vector can be expressed as a linear combination of $\{\vec{\mathcal{F}}_{a},\vec{\mathcal{G}}_{a}\ (a,b=1,2,\cdots,N)\}$.

In terms of $\{\vec{\mathcal{F}}_{a},\vec{\mathcal{G}}_{a}\ (a,b=1,2,\cdots,N)\}$, the boundary vectors {$\vec{F}^{(i)}_{0}$} and {$\vec{G}^{(j)}_{0}$} given in eq.~(\ref{boundary-vectors}) associated with the zero mode solutions (\ref{f0}) {and} (\ref{g0}) can be expressed as follows:
	\al{
	\vec{F}^{(i)}_{0}=\sum^{N}_{a=1}A_{a}^{(i)}\vec{\mathcal{F}}_{a},\label{F-decomposition}\\
	\vec{G}^{(j)}_{0}=\sum^{N}_{a=1}B_{a}^{(j)}\vec{\mathcal{G}}_{a}.\label{G-decomposition}
	}

In the following, we shall {clarify} how many zero mode solutions exist for each of the {type\,($2N-k,k$)} BCs ($k=0,1,2,\cdots,2N$). This is equivalent to find linearly independent boundary vectors $\vec{F}^{(i)}_{0}$ and $\vec{G}^{(j)}_{0}$ which satisfy the boundary conditions (\ref{BC-F}) {and} (\ref{BC-G}) for a given $U$. To this end, it is convenient to use the fact that any $2N\times 2N$ unitary matrix $V$ can be expressed as
	\al{
	V=(\vec{u}_{1},\vec{u}_{2},\cdots,\vec{u}_{2N}),
	} 
where $\vec{u}_{p}$ ($p=1,2,\cdots,2N$) are $2N$-dimensional {orthonormal complex vectors} satisfying $\vec{u}_{p}{}^{\dagger}\vec{u}_{p'}=\delta_{pp'}$ ($p,p'=1,2,\cdots,2N$). Then, the matrix $U$ belonging to the {type\,($2N-k,k$)} BC will be given by
	\al{
	U=\sum^{2N-k}_{p=1}\vec{u}_{p}\vec{u}_{p}^{\dagger}-\sum^{2N}_{q=2N-k+1}\vec{u}_{q}\vec{u}_{q}^{\dagger}.
	}
It follows that the {type\,($2N-k,k$) BCs}, (\ref{BC-F}) {and} (\ref{BC-G}), reduce to
	\al{
	\vec{u}_{q}^{\dagger}\vec{F}_{0}^{(i)}&=0\qquad \text{for}\quad q=2N-k+1, 2N-k+2, \cdots, 2N,	
\label{BC-Fu}\\
	\vec{u}_{p}^{\dagger}\vec{G}_{0}^{(j)}&=0\qquad \text{for}\quad p=1,2,\cdots,2N-k.\label{BC-Gu}
	}

%%%%%%%%%%%%\UTF{0081}	subsection.4.1	%%%%%%%%%%%%%
\subsection{{{Type}\,($2N-k,k$)} BC with $0\leq k\leq N$}
\qquad Let us first consider the case of $0\leq k\leq N$. Since the set of {$\{\vec{\mathcal{F}}_{a},\vec{\mathcal{G}}_{a}\ (a=1,2,\cdots,N)\}$}  forms a complete set of the $2N$-dimensional complex vector space, the vectors {$\{\vec{u}_{q}\, (q=2N-k+1,\cdots,2N)\}$} can be expressed as
	\al{
	\vec{u}_{q}=\sum^{N}_{a=1}\alpha_{q,{a}}\vec{\mathcal{F}}_{a}+\sum^{N}_{a=1}\beta_{q,a}\vec{\mathcal{G}}_{a}\label{2Nvector-decomposition}
	}
for ${}^{\exists}\alpha_{q,a}, \beta_{q,a}\in\mathbb{C}$ ($q=2N-k+1,\cdots,2N$). Inserting eqs.~($\ref{F-decomposition}$) and (\ref{2Nvector-decomposition}) into eq.~(\ref{BC-Fu}) and using eqs.~(\ref{FFGG-normalization}) and (\ref{FFGG-orthogonal-relation}), we have
	\al{
	\mbox{\boldmath $\alpha$}^{\dagger}_{q}\,{\cdot}\,\mbox{\boldmath $A$}^{(i)}=0,\qquad (q=2N-k+1,\cdots,2N){,}\label{alphaA-condition}
	}
where $\mbox{\boldmath $\alpha$}_{q}\equiv (\alpha_{q,1},\alpha_{q,2},\cdots,\alpha_{q,N})^{{\rm T}}$.

If the maximal number of the linearly independent vectors for {$\{\mbox{\boldmath $\alpha$}_{q}\,(q=2N-k+1,\cdots,2N)\}$} is $l$ with $0\leq l\leq k$, it turns out that there exist $(N-l)$ linearly independent solutions of $\mbox{\boldmath $A$}^{(i)}$ $(i=1,2,\cdots,N-l)$ to eq.~(\ref{alphaA-condition}). This immediately implies that there are $(N-l)$ linearly independent boundary vectors $\vec{F}^{(i)}_{0}$ $(i=1,2,\cdots,N-l)$.

In order to obtain $\vec{G}^{(j)}_{0}$, we first notice that {solutions} of $\vec{G}^{(j)}_{0}$ to eq.~(\ref{BC-Gu}) could be written into the form
	\al{
	\vec{G}^{(j)}_{0}=\sum^{2N}_{q=2N-k+1}b^{(j)}_{q}{\vec{u}_{q}}\label{vecG-expansion}
	}
for ${}^{\exists}b^{(j)}_{q}\in\mathbb{C}$ ($q=2N-k+1,\cdots,2N$). Since we have assumed that the set of the vectors {$\{\mbox{\boldmath $\alpha$}_{q}\,(q=2N-k+1,\cdots,2N)\}$ in eq.~(\ref{2Nvector-decomposition})} consists of $l$ linearly independent vectors, by appropriately choosing the coefficients $b^{(j)}_{q}\in\mathbb{C}$ in eq.~(\ref{vecG-expansion}), we can obtain $(k-l)$ linearly independent solutions of the boundary vectors $\vec{G}^{(j)}_{0}$ ($j=1,2,\cdots, k-l$), which are written into the form (\ref{G-decomposition}). 

Therefore, we conclude that for the {type\,($2N-k,k$)} BC with $0\leq k\leq N$, the number of the zero mode solutions $f^{(i)}_{0}(y)$ and $g^{{(j)}}_{0}(y)$ are given as follows:\\[0.1cm]
	%%%%%%%%%%%%%%%%%%%%%%%%%%%%%%%%%
	\begin{table}[H]
		\centering
 	{Type\,($2N-k,k$)} BC with $0\leq k\leq N$\\[0.3cm]
 	\label{table:2N-kBC}
	{\tabcolsep=15mm
  	\begin{tabular}{c|c|c}
   	\hline
	$l$&$N_{f_{0}}$&$N_{g_{0}}$\\
  	\hline 
	
	$0$&$N$&$k$\\
	$1$&$N-1$&$k-1$\\
	\vdots&\vdots&\vdots\\
	$k$&$N-k$&$0$\\
	\hline
	\end{tabular}
	}
	\caption{The number of the zero mode solutions of $f^{(i)}_{0}(y)$ and $g^{(j)}_{0}(y)$ for the {type\,($2N-k,k$)} BC ($0\leq k\leq N$). $l$ denotes the maximal number of the linearly independent vectors {$\{\mbox{\boldmath $\alpha$}_{q}\,(q=2N-k+1,\cdots,2N)\}$} in eq.~(\ref{2Nvector-decomposition}). $N_{f_{0}}$ ($N_{g_{0}}$) is the number of the zero mode solutions of $f^{(i)}_{0}(y)$ ($g^{{(j)}}_{0}(y)$).}
	\end{table}
	%%%%%%%%%%%%%%%%%%%%%%%%%%%%%%%%

%%%%%%%%%%%%\UTF{0081}	subsection.4.2	%%%%%%%%%%%%%
\subsection{{{Type}\,($2N-k,k$)} BC with $N\leq k\leq 2N$}
\qquad Let us next consider the case of $N\leq k\leq 2N$. In this case, we will expand $\vec{u}_{p}$ ($p=1,2,\cdots, 2N-k$) as 
	\al{
	\vec{u}_{p}=\sum^{N}_{a=1}\alpha_{p,{a}}\vec{\mathcal{F}}_{a}+\sum^{N}_{a=1}\beta_{p,a}\vec{\mathcal{G}}_{a}\label{2Nvector-decomposition-2}
	}
for ${}^{\exists}\alpha_{p,a}, \beta_{p,a}\in\mathbb{C}$ ($p=1,2,\cdots,2N-k$).  Inserting eqs.~(\ref{G-decomposition}) and (\ref{2Nvector-decomposition-2}) into eq.~(\ref{BC-Gu}) and using eqs.~(\ref{FFGG-normalization}) and (\ref{FFGG-orthogonal-relation}), we have
	\al{
	{\mbox{\boldmath $\beta$}_{p}^{\dagger}}\,{\cdot}\,\mbox{\boldmath $B$}^{(j)}=0,\qquad (p=1,2,\cdots,2N-k)\label{betaB-condition}
	}
where $\mbox{\boldmath $\beta$}_{p}\equiv (\beta_{p,1},\beta_{p,2},\cdots,\beta_{p,N})^{\rm T}$. If the maximal number of the linearly independent vectors for {$\{\mbox{\boldmath $\beta$}_{p}\,(p=1,2,\cdots,2N-k)\}$} is $l$ with $0\leq l \leq 2N-k$, it turns out that there exist $(N-l)$ linearly independent solutions of $\mbox{\boldmath $B$}^{(j)}$ ($j=1,2,\cdots,N-l$) to eq.~(\ref{betaB-condition}). This implies that there $(N-l)$ linearly independent vectors $\vec{G}^{(j)}_{0}$ ($j=1,2,\cdots,N-l$).

In order to obtain $\vec{F}^{(i)}_{0}$, we first notice that solutions of $\vec{F}^{(i)}_{0}$ to 
eq.~(\ref{BC-Fu}) could be written into the form
	\al{
	\vec{F}^{(i)}_{0}=\sum^{2N-k}_{p=1}a_{p}^{(i)}\vec{u}_{p},
	}
for ${}^{\exists}a^{(i)}_{p}\in\mathbb{C}$ ($p=1,2,\cdots,2N-k$). Since we have assumed that the set of the vectors $\{\mbox{\boldmath $\beta$}_{p} (p=1,2,\cdots,2N-k)\}$ in eq.~(\ref{2Nvector-decomposition}) consists of $l$ linearly independent vectors, by appropriately choosing the coefficients $a^{(i)}_{p}\in\mathbb{C}$, we can obtain $(2N-k-l)$ linearly independent solutions of the boundary vectors $\vec{F}^{(i)}_{0}$ ($i=1,2,\cdots,2N-k-l$), which are written into the form (\ref{F-decomposition}).

Therefore, we conclude that for the {type\,($2N-k,k$)} BC with $N\leq k\leq 2N$, the number of the zero mode solutions $f^{(i)}_{0}(y)$ and $g^{(j)}_{0}(y)$ are given as follows:\\[0.1cm]
	%%%%%%%%%%%%%%%%%%%%%%%%%%%%%%%%%
	\begin{table}[H]
		\centering
 	{Type\,($2N-k,k$)} BC with $N\leq k\leq 2N$\\[0.3cm]
 	\label{table:2N-kBC-2}
	{\tabcolsep=15mm
  	\begin{tabular}{c|c|c}
   	\hline
	$l$&$N_{f_{0}}$&$N_{g_{0}}$\\
  	\hline 
	
	$0$&$2N-k$&$N$\\
	$1$&$2N-k-1$&$N-1$\\
	\vdots&\vdots&\vdots\\
	$2N-k$&$0$&$-N+k$\\
	\hline
	\end{tabular}
	}
	\caption{The number of the zero mode solutions of $f^{(i)}_{0}(y)$ and $g^{(j)}_{0}(y)$ for the {type\,($2N-k,k$)} BC ($N\leq k\leq 2N$). $l$ denotes the maximal number of the linearly independent vectors {$\{\mbox{\boldmath $\beta$}_{p}\,(p=1,2,\cdots,2N-k)\}$} in eq.~(\ref{2Nvector-decomposition-2}). $N_{f_{0}}$ ($N_{g_{0}}$) is the number of the zero mode solutions of $f^{(i)}_{0}(y)$ ($g^{{(j)}}_{0}(y)$).}
	\end{table}
	%%%%%%%%%%%%%%%%%%%%%%%%%%%%%%%%

%%%%%%%%%%%%\UTF{0081}	section.5	%%%%%%%%%%%%%
\section{Zero mode solutions and Witten index}
\qquad In the previous section, we have succeeded in finding the number $N_{f_{0}}$ ($N_{g_{0}}$) of the zero mode solutions $f^{(i)}_{0}(y)$ ($g^{{(j)}}_{0}(y)$). Then, we have seen that $N_{f_{0}}$ and $N_{g_{0}}$ depend on $l$ and $k$, where $l$ is the maximal number of linearly independent vectors for {$\{\mbox{\boldmath $\alpha$}_{q}\,(q=2N-k+1,\cdots,2N)\}$} in eq.~(\ref{2Nvector-decomposition}) or {$\{\mbox{\boldmath $\beta$}_{p}\,(p=1,2,\cdots,2N-k)\}$} in eq.~(\ref{2Nvector-decomposition-2}), and $k$ is the number that specifies the {type\,($2N-k,k$)} BC. It will be worthwhile noting that the difference $\Delta_{\rm W}\equiv N_{f_{0}}-N_{g_{0}}$ is independent of {$l$}, {\textit{i.e.}} $\Delta_{\rm W}=N-k$ for the {type\,($2N-k,k$)} BC with $0\leq k\leq 2N$, though $N_{f_{0}}$ and $N_{g_{0}}$ depend on $l$ {(see {Table 1 and Table 2})}. The fact that $\Delta_{\rm W}$ is independent of $l$ is not accidental, but it has a topological reason, as we will explain below.

In the model of the 5d Dirac action on an extra dimension, a {quantum}-mechanical supersymmetry has been shown to be hidden in the 4d mass spectrum~\cite{Fujimoto:2012wv}. The Hermitian operator $H$, $Q$ and $(-1)^{F}$ defined by
	\al{
	H=Q^{2},\quad Q=\mat{0&-\partial_{y}+M\\ \partial_{y}+M&0},\quad (-1)^{F}=\mat{1&0\\0&-1},
	}
are found to form a supersymmetric quantum mechanics{, where $H$ is the Hamiltonian, $Q$ denotes the supercharge and $(-1)^{F}$ is called the fermion number operator}. Then, the Witten index defined by
	\al{
	\Delta_{\rm W}\equiv N^{E=0}_{(-1)^{F}=+1}-N^{E=0}_{(-1)^{F}=-1}
	}
is known to be a topological index. Here, $N^{E=0}_{(-1)^{F}=\pm 1}$ denote the {numbers} of the solutions with $H=0$ and $(-1)^{F}=\pm 1$, respectively. A non-trivial fact is that the supercharge $Q$ is Hermitian on the rose graph depicted in {fig.~\ref{fig:rose_graph}} with the boundary conditions (\ref{BC-F}) and (\ref{BC-G}). The relations (\ref{SUSYQM-f}) and (\ref{SUSYQM-g}) for the mode functions $f^{(i)}_{n}(y)$ and $g^{(i)}_{n}(y)$ {imply that they form supermultiplets for $m_{n}\neq 0$}.

Since the {type\,($2N-k,k$)} BC is not continuously connected to other ($2N-k',k'$) BC with $k\neq k'$, the Witten index $\Delta_{\rm W}$ can depend on $k$ but not on $l$ because the boundary conditions with different $l$ can continuously be deformed each other.

%%%%%%%%%%%%\UTF{0081}	section.6	%%%%%%%%%%%%%
\section{Conclusion and Discussion}
\quad In this paper, we have investigated {the KK decomposition of the 5d Dirac fermion} on the rose graph which consists of one vertex and {$N$} loops (see {fig.~\ref{fig:rose_graph}}). We have succeeded in classifying the allowed boundary condition on the rose graph into the {type\,($2N-k,k$)} BC with $0\leq k\leq 2N$ and finding the number of the zero mode solutions for each type of {the} boundary conditions.

Our results would become phenomenologically important in constructing models beyond the standard model, based on our model considered in this paper. The chiral fermions of the standard model may correspond to {the} 4d chiral massless fermions {$\psi^{(i)}_{{\rm R},0}(x)$ and $\psi^{(j)}_{{\rm L},0}(x)$} associated with the zero mode solutions $f^{(i)}_{0}(y)$ and $g^{(j)}_{0}(y)$. Thus, the three generation of the quarks and leptons could be obtained from a model with ${|\Delta_{\rm W}|}=3$, {\textit{i.e.}} the {type\,($N-3,N+3$)} BC or {type\,($N+3,N-3$)} BC for the rose graph with $N$ loops. The fermion mass hierarchy problem in the standard model will be solved in our model because the zero mode solutions $f^{(i)}_{0}(y)$ and $g^{(j)}_{0}(y)$ are {exponentially} localized at some boundaries (see eqs.~(\ref{f0}) and (\ref{g0})) and the overlap {integrals} of zero mode solutions would produce hierarchical masses for quarks and leptons. Furthermore, our model has a natural source of a \textit{CP} violating phase in the CKM matrix. This is because the boundary conditions (\ref{BC-F}) and (\ref{BC-G}) include complex parameters, in general, through the {complex (Hermitian)} matrix $U$, {so that} the zero mode solutions $f^{(i)}_{0}(y)$ and $g^{(j)}_{0}(y)$ could become genuine complex functions. Therefore, our model considered in this paper is expected to shed a new light on the generation problem, the fermion mass hierarchy and the \textit{CP}-violating phase of the standard model. Phenomenological applications of our model will be reported elsewhere.

It would be of great interest to note that the rose graph depicted in {fig.~\ref{fig:rose_graph}} has non-trivial geometry and that the parameter space of the boundary conditions has been found to possess a rich structure. For instance, if we take the length of all bonds to be equal ({$L= L_{a}-L_{a-1}$} for $a=1,2,\cdots, N$) and choose the boundary conditions appropriately, we will have higher degeneracy in the 4d mass spectrum and then extended supersymmetries might appear in the spectrum~{\cite{Nagasawa:2003tw,Nagasawa:2005kv}}. Furthermore, we will expect non-abelian Berry phases in the space of the boundary conditions with the degeneracy of the spectrum~{\cite{Wilczek:1984dh,Ohya:2014ska,Ohya:2015xya}}. The issues mentioned above will be discussed in a forthcoming paper.

%%%%%%%%%%\UTF{0081}%%%%%%%%%%%%%	
\section*{Acknowledgements}
	{This work was supported by JSPS KAKENHI Grant Number JP 18K03649~(Y.F., M.S. and K.T.)}. {{The authors} thank K.~Hasegawa and P.~Tanaka for useful {discussions}.}
	
\bibliographystyle{utphys}
\bibliography{references,references_2}

\providecommand{\href}[2]{#2}\begingroup\raggedright\begin{thebibliography}{10}

\bibitem{Libanov:2000uf}
M.~V. Libanov and S.~V. Troitsky, ``{Three fermionic generations on a
  topological defect in extra dimensions},''
  \href{http://dx.doi.org/10.1016/S0550-3213(01)00036-0}{{\em Nucl. Phys.}
  {\bfseries B599} (2001) 319--333},
\href{http://arxiv.org/abs/hep-ph/0011095}{{\ttfamily arXiv:hep-ph/0011095
  [hep-ph]}}.
%%CITATION = HEP-PH/0011095;%%.

\bibitem{Frere:2000dc}
J.~M. Frere, M.~V. Libanov, and S.~V. Troitsky, ``{Three generations on a local
  vortex in extra dimensions},''
  \href{http://dx.doi.org/10.1016/S0370-2693(01)00696-7}{{\em Phys. Lett.}
  {\bfseries B512} (2001) 169--173},
\href{http://arxiv.org/abs/hep-ph/0012306}{{\ttfamily arXiv:hep-ph/0012306
  [hep-ph]}}.
%%CITATION = HEP-PH/0012306;%%.

\bibitem{Blumenhagen:2000wh}
R.~Blumenhagen, L.~Goerlich, B.~Kors, and D.~Lust, ``{Noncommutative
  compactifications of type I strings on tori with magnetic background flux},''
  \href{http://dx.doi.org/10.1088/1126-6708/2000/10/006}{{\em JHEP} {\bfseries
  10} (2000) 006},
\href{http://arxiv.org/abs/hep-th/0007024}{{\ttfamily arXiv:hep-th/0007024
  [hep-th]}}.
%%CITATION = HEP-TH/0007024;%%.

\bibitem{Blumenhagen:2000ea}
R.~Blumenhagen, B.~Kors, and D.~Lust, ``{Type I strings with F flux and B
  flux},'' \href{http://dx.doi.org/10.1088/1126-6708/2001/02/030}{{\em JHEP}
  {\bfseries 02} (2001) 030},
\href{http://arxiv.org/abs/hep-th/0012156}{{\ttfamily arXiv:hep-th/0012156
  [hep-th]}}.
%%CITATION = HEP-TH/0012156;%%.

\bibitem{Cremades:2004wa}
D.~Cremades, L.~E. Ibanez, and F.~Marchesano, ``{Computing Yukawa couplings
  from magnetized extra dimensions},''
  \href{http://dx.doi.org/10.1088/1126-6708/2004/05/079}{{\em JHEP} {\bfseries
  05} (2004) 079},
\href{http://arxiv.org/abs/hep-th/0404229}{{\ttfamily arXiv:hep-th/0404229
  [hep-th]}}.
%%CITATION = HEP-TH/0404229;%%.

\bibitem{Abe:2008fi}
H.~Abe, T.~Kobayashi, and H.~Ohki, ``{Magnetized orbifold models},''
  \href{http://dx.doi.org/10.1088/1126-6708/2008/09/043}{{\em JHEP} {\bfseries
  09} (2008) 043},
\href{http://arxiv.org/abs/0806.4748}{{\ttfamily arXiv:0806.4748 [hep-th]}}.
%%CITATION = ARXIV:0806.4748;%%.

\bibitem{Abe:2008sx}
H.~Abe, K.-S. Choi, T.~Kobayashi, and H.~Ohki, ``{Three generation magnetized
  orbifold models},''
  \href{http://dx.doi.org/10.1016/j.nuclphysb.2009.02.002}{{\em Nucl. Phys.}
  {\bfseries B814} (2009) 265--292},
\href{http://arxiv.org/abs/0812.3534}{{\ttfamily arXiv:0812.3534 [hep-th]}}.
%%CITATION = ARXIV:0812.3534;%%.

\bibitem{Abe:2013bca}
T.-H. Abe, Y.~Fujimoto, T.~Kobayashi, T.~Miura, K.~Nishiwaki, and M.~Sakamoto,
  ``{$Z_N$ twisted orbifold models with magnetic flux},''
  \href{http://dx.doi.org/10.1007/JHEP01(2014)065}{{\em JHEP} {\bfseries 01}
  (2014) 065},
\href{http://arxiv.org/abs/1309.4925}{{\ttfamily arXiv:1309.4925 [hep-th]}}.
%%CITATION = ARXIV:1309.4925;%%.

\bibitem{Fujimoto:2013xha}
Y.~Fujimoto, T.~Kobayashi, T.~Miura, K.~Nishiwaki, and M.~Sakamoto, ``{Shifted
  orbifold models with magnetic flux},''
  \href{http://dx.doi.org/10.1103/PhysRevD.87.086001}{{\em Phys. Rev.}
  {\bfseries D87} no.~8, (2013) 086001},
\href{http://arxiv.org/abs/1302.5768}{{\ttfamily arXiv:1302.5768 [hep-th]}}.
%%CITATION = ARXIV:1302.5768;%%.

\bibitem{Abe:2014noa}
T.-h. Abe, Y.~Fujimoto, T.~Kobayashi, T.~Miura, K.~Nishiwaki, and M.~Sakamoto,
  ``{Operator analysis of physical states on magnetized $T^{2}/Z_{N}$
  orbifolds},'' \href{http://dx.doi.org/10.1016/j.nuclphysb.2014.11.022}{{\em
  Nucl. Phys.} {\bfseries B890} (2014) 442--480},
\href{http://arxiv.org/abs/1409.5421}{{\ttfamily arXiv:1409.5421 [hep-th]}}.
%%CITATION = ARXIV:1409.5421;%%.

\bibitem{Abe:2015yva}
T.-h. Abe, Y.~Fujimoto, T.~Kobayashi, T.~Miura, K.~Nishiwaki, M.~Sakamoto, and
  Y.~Tatsuta, ``{Classification of three-generation models on magnetized
  orbifolds},'' \href{http://dx.doi.org/10.1016/j.nuclphysb.2015.03.004}{{\em
  Nucl. Phys.} {\bfseries B894} (2015) 374--406},
\href{http://arxiv.org/abs/1501.02787}{{\ttfamily arXiv:1501.02787 [hep-ph]}}.
%%CITATION = ARXIV:1501.02787;%%.

\bibitem{Fujimoto:2016llj}
Y.~Fujimoto, K.~Hasegawa, K.~Nishiwaki, M.~Sakamoto, and K.~Tatsumi, ``{6d
  Dirac fermion on a rectangle; scrutinizing boundary conditions, mode
  functions and spectrum},''
  \href{http://dx.doi.org/10.1016/j.nuclphysb.2017.06.024}{{\em Nucl. Phys.}
  {\bfseries B922} (2017) 186--225},
\href{http://arxiv.org/abs/1609.01413}{{\ttfamily arXiv:1609.01413 [hep-th]}}.
%%CITATION = ARXIV:1609.01413;%%.

\bibitem{Fujimoto:2016rbr}
Y.~Fujimoto, K.~Hasegawa, K.~Nishiwaki, M.~Sakamoto, and K.~Tatsumi,
  ``{Supersymmetry in the 6D Dirac action},''
  \href{http://dx.doi.org/10.1093/ptep/ptx092}{{\em PTEP} {\bfseries 2017}
  no.~7, (2017) 073B03},
\href{http://arxiv.org/abs/1609.04565}{{\ttfamily arXiv:1609.04565 [hep-th]}}.
%%CITATION = ARXIV:1609.04565;%%.

\bibitem{Fujimoto:2018cnf}
Y.~Fujimoto, K.~Hasegawa, K.~Nishiwaki, M.~Sakamoto, K.~Tatsumi, and I.~Ueba,
  ``{Extended supersymmetry in Dirac action with extra dimensions},''
  \href{http://dx.doi.org/10.1088/1751-8121/aadea2}{{\em J. Phys.} {\bfseries
  A51} no.~43, (2018) 435201},
\href{http://arxiv.org/abs/1804.02626}{{\ttfamily arXiv:1804.02626 [hep-th]}}.
%%CITATION = ARXIV:1804.02626;%%.

\bibitem{Fujimoto:2018tjm}
Y.~Fujimoto, K.~Hasegawa, K.~Nishiwaki, M.~Sakamoto, K.~Tatsumi, and I.~Ueba,
  ``{Extended supersymmetry with central charges in higher dimensional Dirac
  action},'' \href{http://dx.doi.org/10.1103/PhysRevD.99.065002}{{\em Phys.
  Rev.} {\bfseries D99} no.~6, (2019) 065002},
\href{http://arxiv.org/abs/1812.11282}{{\ttfamily arXiv:1812.11282 [hep-th]}}.
%%CITATION = ARXIV:1812.11282;%%.

\bibitem{ArkaniHamed:1998rs}
N.~Arkani-Hamed, S.~Dimopoulos, and G.~R. Dvali, ``{The Hierarchy problem and
  new dimensions at a millimeter},''
  \href{http://dx.doi.org/10.1016/S0370-2693(98)00466-3}{{\em Phys. Lett.}
  {\bfseries B429} (1998) 263--272},
\href{http://arxiv.org/abs/hep-ph/9803315}{{\ttfamily arXiv:hep-ph/9803315
  [hep-ph]}}.
%%CITATION = HEP-PH/9803315;%%.

\bibitem{ArkaniHamed:1999dc}
N.~Arkani-Hamed and M.~Schmaltz, ``{Hierarchies without symmetries from extra
  dimensions},'' \href{http://dx.doi.org/10.1103/PhysRevD.61.033005}{{\em Phys.
  Rev.} {\bfseries D61} (2000) 033005},
\href{http://arxiv.org/abs/hep-ph/9903417}{{\ttfamily arXiv:hep-ph/9903417
  [hep-ph]}}.
%%CITATION = HEP-PH/9903417;%%.

\bibitem{Dvali:2000ha}
G.~R. Dvali and M.~A. Shifman, ``{Families as neighbors in extra dimension},''
  \href{http://dx.doi.org/10.1016/S0370-2693(00)00083-6}{{\em Phys. Lett.}
  {\bfseries B475} (2000) 295--302},
\href{http://arxiv.org/abs/hep-ph/0001072}{{\ttfamily arXiv:hep-ph/0001072
  [hep-ph]}}.
%%CITATION = HEP-PH/0001072;%%.

\bibitem{Gherghetta:2000qt}
T.~Gherghetta and A.~Pomarol, ``{Bulk fields and supersymmetry in a slice of
  AdS},'' \href{http://dx.doi.org/10.1016/S0550-3213(00)00392-8}{{\em Nucl.
  Phys.} {\bfseries B586} (2000) 141--162},
\href{http://arxiv.org/abs/hep-ph/0003129}{{\ttfamily arXiv:hep-ph/0003129
  [hep-ph]}}.
%%CITATION = HEP-PH/0003129;%%.

\bibitem{Huber:2000ie}
S.~J. Huber and Q.~Shafi, ``{Fermion masses, mixings and proton decay in a
  Randall-Sundrum model},''
  \href{http://dx.doi.org/10.1016/S0370-2693(00)01399-X}{{\em Phys. Lett.}
  {\bfseries B498} (2001) 256--262},
\href{http://arxiv.org/abs/hep-ph/0010195}{{\ttfamily arXiv:hep-ph/0010195
  [hep-ph]}}.
%%CITATION = HEP-PH/0010195;%%.

\bibitem{Kaplan:2001ga}
D.~E. Kaplan and T.~M.~P. Tait, ``{New tools for fermion masses from extra
  dimensions},'' \href{http://dx.doi.org/10.1088/1126-6708/2001/11/051}{{\em
  JHEP} {\bfseries 11} (2001) 051},
\href{http://arxiv.org/abs/hep-ph/0110126}{{\ttfamily arXiv:hep-ph/0110126
  [hep-ph]}}.
%%CITATION = HEP-PH/0110126;%%.

\bibitem{Fujimoto:2011kf}
Y.~Fujimoto, T.~Nagasawa, S.~Ohya, and M.~Sakamoto, ``{Phase Structure of Gauge
  Theories on an Interval},'' \href{http://dx.doi.org/10.1143/PTP.126.841}{{\em
  Prog. Theor. Phys.} {\bfseries 126} (2011) 841--854},
\href{http://arxiv.org/abs/1108.1976}{{\ttfamily arXiv:1108.1976 [hep-th]}}.
%%CITATION = ARXIV:1108.1976;%%.

\bibitem{Fujimoto:2013ki}
Y.~Fujimoto, K.~Nishiwaki, and M.~Sakamoto, ``{\textit{CP} phase from twisted
  Higgs vacuum expectation value in extra dimension},''
  \href{http://dx.doi.org/10.1103/PhysRevD.88.115007}{{\em Phys. Rev.}
  {\bfseries D88} no.~11, (2013) 115007},
\href{http://arxiv.org/abs/1301.7253}{{\ttfamily arXiv:1301.7253 [hep-ph]}}.
%%CITATION = ARXIV:1301.7253;%%.

\bibitem{Kuchment_2004}
P.~Kuchment, ``Quantum graphs: I. some basic structures,''
  \href{http://dx.doi.org/10.1088/0959-7174/14/1/014}{{\em Waves in Random
  Media} {\bfseries 14} no.~1, (2004) S107--S128}.

\bibitem{Kuchment_2005}
P.~Kuchment, ``{Quantum graphs: II. Some spectral properties of quantum and
  combinatorial graphs},''
  \href{http://dx.doi.org/10.1088/0305-4470/38/22/013}{{\em Journal of Physics
  A: Mathematical and General} {\bfseries 38} no.~22, (2005) 4887--4900},
  \href{http://arxiv.org/abs/math-ph/0411003}{{\ttfamily arxiv:math-ph/0411003
  [math-ph]}}.

\bibitem{Berkolaiko_1999}
G.~Berkolaiko and J.~P. Keating, ``Two-point spectral correlations for star
  graphs,'' \href{http://dx.doi.org/10.1088/0305-4470/32/45/302}{{\em Journal
  of Physics A: Mathematical and General} {\bfseries 32} no.~45, (1999)
  7827--7841}.

\bibitem{Bellazzini:2008mn}
B.~Bellazzini, M.~Burrello, M.~Mintchev, and P.~Sorba, ``{Quantum Field Theory
  on Star Graphs},'' {\em Proc. Symp. Pure Math.} {\bfseries 77} (2008) 639,
\href{http://arxiv.org/abs/0801.2852}{{\ttfamily arXiv:0801.2852 [hep-th]}}.
%%CITATION = ARXIV:0801.2852;%%.

\bibitem{Adami:2010zj}
R.~Adami, C.~Cacciapuoti, D.~Finco, and D.~Noja, ``{Fast solitons on star
  graphs},'' \href{http://dx.doi.org/10.1142/S0129055X11004345}{{\em Rev. Math.
  Phys.} {\bfseries 23} (2011) 409--451},
\href{http://arxiv.org/abs/1004.2455}{{\ttfamily arXiv:1004.2455 [math-ph]}}.
%%CITATION = ARXIV:1004.2455;%%.

\bibitem{Fujimoto:2018lzq}
Y.~Fujimoto, K.~Konno, T.~Nagasawa, and R.~Takahashi, ``{Quantum Reflection and
  Transmission in Ring Systems with Double Y-Junctions: Occurrence of Perfect
  Reflection},''
\href{http://arxiv.org/abs/1812.05749}{{\ttfamily arXiv:1812.05749
  [quant-ph]}}.
%%CITATION = ARXIV:1812.05749;%%.

\bibitem{Dimopoulos:2001qd}
S.~Dimopoulos, S.~Kachru, N.~Kaloper, A.~E. Lawrence, and E.~Silverstein,
  ``{Generating small numbers by tunneling in multithroat compactifications},''
  \href{http://dx.doi.org/10.1142/S0217751X04018075}{{\em Int. J. Mod. Phys.}
  {\bfseries A19} (2004) 2657--2704},
\href{http://arxiv.org/abs/hep-th/0106128}{{\ttfamily arXiv:hep-th/0106128
  [hep-th]}}.
%%CITATION = HEP-TH/0106128;%%.

\bibitem{Kim:2005aa}
H.~D. Kim, ``{Hiding an extra dimension},''
  \href{http://dx.doi.org/10.1088/1126-6708/2006/01/090}{{\em JHEP} {\bfseries
  01} (2006) 090},
\href{http://arxiv.org/abs/hep-th/0510229}{{\ttfamily arXiv:hep-th/0510229
  [hep-th]}}.
%%CITATION = HEP-TH/0510229;%%.

\bibitem{Cacciapaglia:2006tg}
G.~Cacciapaglia, C.~Csaki, C.~Grojean, and J.~Terning, ``{Field Theory on
  Multi-throat Backgrounds},''
  \href{http://dx.doi.org/10.1103/PhysRevD.74.045019}{{\em Phys. Rev.}
  {\bfseries D74} (2006) 045019},
\href{http://arxiv.org/abs/hep-ph/0604218}{{\ttfamily arXiv:hep-ph/0604218
  [hep-ph]}}.
%%CITATION = HEP-PH/0604218;%%.

\bibitem{Bechinger:2009qk}
A.~Bechinger and G.~Seidl, ``{Resonant Dirac leptogenesis on throats},''
  \href{http://dx.doi.org/10.1103/PhysRevD.81.065015}{{\em Phys. Rev.}
  {\bfseries D81} (2010) 065015},
\href{http://arxiv.org/abs/0907.4341}{{\ttfamily arXiv:0907.4341 [hep-ph]}}.
%%CITATION = ARXIV:0907.4341;%%.

\bibitem{Abel:2010kw}
S.~Abel and J.~Barnard, ``{Strong coupling, discrete symmetry and flavour},''
  \href{http://dx.doi.org/10.1007/JHEP08(2010)039}{{\em JHEP} {\bfseries 08}
  (2010) 039},
\href{http://arxiv.org/abs/1005.1668}{{\ttfamily arXiv:1005.1668 [hep-ph]}}.
%%CITATION = ARXIV:1005.1668;%%.

\bibitem{Law:2010pv}
S.~S.~C. Law and K.~L. McDonald, ``{Broken Symmetry as a Stabilizing
  Remnant},'' \href{http://dx.doi.org/10.1103/PhysRevD.82.104032}{{\em Phys.
  Rev.} {\bfseries D82} (2010) 104032},
\href{http://arxiv.org/abs/1008.4336}{{\ttfamily arXiv:1008.4336 [hep-ph]}}.
%%CITATION = ARXIV:1008.4336;%%.

\bibitem{Nagasawa:2002un}
T.~Nagasawa, M.~Sakamoto, and K.~Takenaga, ``{Supersymmetry in quantum
  mechanics with point interactions},''
  \href{http://dx.doi.org/10.1016/S0370-2693(03)00575-6}{{\em Phys. Lett.}
  {\bfseries B562} (2003) 358--364},
\href{http://arxiv.org/abs/hep-th/0212192}{{\ttfamily arXiv:hep-th/0212192
  [hep-th]}}.
%%CITATION = HEP-TH/0212192;%%.

\bibitem{Nagasawa:2003tw}
T.~Nagasawa, M.~Sakamoto, and K.~Takenaga, ``{Supersymmetry and discrete
  transformations on $S^1$ with point singularities},''
  \href{http://dx.doi.org/10.1016/j.physletb.2003.12.065}{{\em Phys. Lett.}
  {\bfseries B583} (2004) 357--363},
\href{http://arxiv.org/abs/hep-th/0311043}{{\ttfamily arXiv:hep-th/0311043
  [hep-th]}}.
%%CITATION = HEP-TH/0311043;%%.

\bibitem{Nagasawa:2005kv}
T.~Nagasawa, M.~Sakamoto, and K.~Takenaga, ``{Extended supersymmetry and its
  reduction on a circle with point singularities},''
  \href{http://dx.doi.org/10.1088/0305-4470/38/37/009}{{\em J. Phys.}
  {\bfseries A38} (2005) 8053--8082},
\href{http://arxiv.org/abs/hep-th/0505132}{{\ttfamily arXiv:hep-th/0505132
  [hep-th]}}.
%%CITATION = HEP-TH/0505132;%%.

\bibitem{Fujimoto:2012wv}
Y.~Fujimoto, T.~Nagasawa, K.~Nishiwaki, and M.~Sakamoto, ``{Quark mass
  hierarchy and mixing via geometry of extra dimension with point
  interactions},'' \href{http://dx.doi.org/10.1093/ptep/pts097}{{\em PTEP}
  {\bfseries 2013} (2013) 023B07},
\href{http://arxiv.org/abs/1209.5150}{{\ttfamily arXiv:1209.5150 [hep-ph]}}.
%%CITATION = ARXIV:1209.5150;%%.

\bibitem{Fujimoto:2014fka}
Y.~Fujimoto, K.~Nishiwaki, M.~Sakamoto, and R.~Takahashi, ``{Realization of
  lepton masses and mixing angles from point interactions in an extra
  dimension},'' \href{http://dx.doi.org/10.1007/JHEP10(2014)191}{{\em JHEP}
  {\bfseries 10} (2014) 191},
\href{http://arxiv.org/abs/1405.5872}{{\ttfamily arXiv:1405.5872 [hep-ph]}}.
%%CITATION = ARXIV:1405.5872;%%.

\bibitem{Fujimoto:2017lln}
Y.~Fujimoto, T.~Miura, K.~Nishiwaki, and M.~Sakamoto, ``{Dynamical generation
  of fermion mass hierarchy in an extra dimension},''
  \href{http://dx.doi.org/10.1103/PhysRevD.97.115039}{{\em Phys. Rev.}
  {\bfseries D97} no.~11, (2018) 115039},
\href{http://arxiv.org/abs/1709.05693}{{\ttfamily arXiv:1709.05693 [hep-th]}}.
%%CITATION = ARXIV:1709.05693;%%.

\bibitem{Kostrykin:1998gz}
V.~Kostrykin and R.~Schrader, ``{Kirchoff's rule for quantum wires},''
\href{http://dx.doi.org/10.1088/0305-4470/32/4/006}{{\em J. Phys.} {\bfseries
  A32} (1999) 595--630}.
%%CITATION = JPAGA,A32,595;%%.

\bibitem{Fulop:1999pf}
T.~Fulop and I.~Tsutsui, ``{A Free particle on a circle with point
  interaction},'' \href{http://dx.doi.org/10.1016/S0375-9601(99)00850-6}{{\em
  Phys. Lett.} {\bfseries A264} (2000) 366},
\href{http://arxiv.org/abs/quant-ph/9910062}{{\ttfamily arXiv:quant-ph/9910062
  [quant-ph]}}.
%%CITATION = QUANT-PH/9910062;%%.

\bibitem{Wilczek:1984dh}
F.~Wilczek and A.~Zee, ``{Appearance of Gauge Structure in Simple Dynamical
  Systems},''
\href{http://dx.doi.org/10.1103/PhysRevLett.52.2111}{{\em Phys. Rev. Lett.}
  {\bfseries 52} (1984) 2111--2114}.
%%CITATION = PRLTA,52,2111;%%.

\bibitem{Ohya:2014ska}
S.~Ohya, ``{Non-Abelian Monopole in the Parameter Space of Point-like
  Interactions},'' \href{http://dx.doi.org/10.1016/j.aop.2014.10.013}{{\em
  Annals Phys.} {\bfseries 351} (2014) 900--913},
\href{http://arxiv.org/abs/1406.4857}{{\ttfamily arXiv:1406.4857 [hep-th]}}.
%%CITATION = ARXIV:1406.4857;%%.

\bibitem{Ohya:2015xya}
S.~Ohya, ``{BPS Monopole in the Space of Boundary Conditions},''
  \href{http://dx.doi.org/10.1088/1751-8113/48/50/505401}{{\em J. Phys.}
  {\bfseries A48} no.~50, (2015) 505401},
\href{http://arxiv.org/abs/1506.04738}{{\ttfamily arXiv:1506.04738 [hep-th]}}.
%%CITATION = ARXIV:1506.04738;%%.

\end{thebibliography}\endgroup

\end{document}